\documentclass[fleqn,usenatbib]{mnras}

\usepackage{newtxtext,newtxmath}
\usepackage[T1]{fontenc}
\usepackage{ae,aecompl}

\usepackage{subcaption}

\usepackage{xcolor}
\usepackage{graphicx}	% Including figure files
\usepackage{amsmath}	% Advanced maths commands

\usepackage{epsfig}
\usepackage{rotating}
\usepackage{color}
\usepackage{tabularx}
\usepackage{multirow}
\usepackage{graphics}
\usepackage{epsfig}
\usepackage{epstopdf}
\usepackage[para,online,flushleft]{threeparttable}
\usepackage[ampersand]{easylist}
\definecolor{navy}{rgb}{0,0.3,0.7}
\definecolor{forrest}{rgb}{0,0.55,0.25}
\definecolor{tyrianpurple}{rgb}{0.4, 0.01, 0.24}
\definecolor{coralred}{rgb}{1.0, 0.25, 0.25}
\usepackage{hyperref}
\hypersetup{
    colorlinks=true,
    linkcolor=blue,
    filecolor=blue,      
    urlcolor=blue,
    citecolor=blue,
    filecolor=blue,
}
\title[Optical emission from G107.5$-$1.5]{Discovery of optical emission associated with the supernova remnant G107.5$-$1.5}

\author[Bak{\i}\c{s} et al.]{{\color {black} H.~Bak{\i}\c{s}$^{1}$,\thanks{E-mail:{\color  {blue}hicranbakis@akdeniz.edu.tr} (HB)}
G.~Bulut$^{1}$,
V.~Bak{\i}\c{s}$^{1}$, H.~Sano$^{2}$ and A.~Sezer$^{3}$}\\
$^{1}$Department of Space Sciences and Technologies, Akdeniz University, 07058, Antalya, Turkey\\
$^{2}$Faculty of Engineering, Gifu University, 1-1 Yanagido, Gifu 501-1193, Japan\\
$^{3}$Department of Electrical-Electronics Engineering, Avrasya University, 61250, Trabzon, Turkey\\
}

\date{Accepted XXX. Received YYY; in original form ZZZ}

\pubyear{2023}

\begin{document}
\label{firstpage}
\pagerange{\pageref{firstpage}--\pageref{lastpage}}
\maketitle

% Abstract of the paper
\begin{abstract}
We present the first results from an imaging and a spectroscopic survey of the optical emission associated with supernova remnant (SNR) G107.5$-$1.5. We discovered optical diffuse and filamentary emission from G107.5$-$1.5 using the 1.5-m and 1-m telescopes. The optical emissions from the North East (NE) and North West (NW) regions show the diffuse structure, while the South East (SE) and East (E) regions show filamentary structure. From long-slit spectra, we found [S\,{\sc ii}]/H$\alpha$ $>$ 0.5 for the SE and E regions, which supports the origin of the emission being from shock-heated gas. The average [S\,{\sc ii}]/H$\alpha$ ratio is found to be $\sim$0.4 and $\sim$0.3 for the NW and NE regions, respectively, indicating that the optical emission is coming from ionized gas in an H\,{\sc ii} region. From the ratios of [S\,{\sc ii}]$\lambda\lambda$ 6716/6731, we estimate an average electron density of $\sim$2400 cm$^{-3}$ for the NW region, which can be attributed to the dense ionized gas. The average [S\,{\sc ii}]$\lambda\lambda$ 6716/6731 ratios are $\sim$1.25 and $\sim$1.15 for the SE and E regions, respectively, which are indicative of low electron density.

\end{abstract}


%%%%%%%%%%%%%%%%%%%%%%%%%%%%%%%%%%%%%%%%%%%%%%%%%%

%%%%%%%%%%%%%%%%% BODY OF PAPER %%%%%%%%%%%%%%%%%%

\section{Introduction}
%% aim \& structure
Observations of supernova remnants (SNRs) provide detailed information about their progenitor star, properties of a supernova (SN) explosion, and physics of shocks as well as the structure of an ambient medium. Most Galactic SNRs have been discovered via radio continuum surveys through the synchrotron emission of non-thermal electrons \citep{Du15}. Currently, approximately 300 confirmed Galactic SNRs have been cataloged \citep{FeSa12, Gr19}. Several SNRs have also been detected first in X-ray surveys.  Only a handful of SNRs have been discovered in the optical band based on CCD imaging (e.g. \citealt{St07, Ma09, Bo09, St09, Fe10, St11, Se12, Fe19, Fe20}). Optical searches for SNRs are mainly done by using an emission line ratio criterion of the form [S\,{\sc ii}]/H$\alpha$ $>$ 0.5 \citep{MaCl73, Fe84, MaFe97, BlLo97, Do10}. 

We systematically search for optical emission from Galactic SNRs using the 1.5-m Russian–Turkish telescope (RTT 150) and 1-m telescope (T100) to investigate the SN explosion mechanism, the shock wave physics of SNRs, and the local ambient medium properties. As part of our systematic search, we discovered optical diffuse and filamentary emission from G107.5$-$1.5.

The shell-type SNR G107.5$-$1.5 was discovered by \cite{Kothes03}  using the Canadian Galactic Plane Survey\footnote{\url{https://ui.adsabs.harvard.edu/abs/2003AJ....125.3145T/abstract}}(CGPS; \citealt{Ta03}) data. The SNR is located in a complex region of the Galactic plane $\sim$4 degree west of Cas A \citep{Kothes03}. 408 MHz and 1420 MHz continuum emission implies a spectral index of $\alpha$ = - 0.6 $\pm$ 0.1 ($S \sim \nu^{\alpha}$), typical for a shell-type SNR. \citet{Kothes03} derived a distance of $d$ $\sim$ 1.1 kpc for G107.5$-$1.5 based on the low rotation measure and possibly related H\,{\sc i} features. \citet{Jackson14} analyzed {\it XMM–Newton} data of G107.5$-$1.5 and did not detect diffuse X-ray emission from the SNR. The authors found eight bright point sources in the field near the centre of the shell. 

%% aim \& structure
In this paper, we present the first results from an imaging and a spectroscopic survey of the optical emission associated with the SNR G107.5$-$1.5. Using the RTT 150 and T100 telescopes, we investigate the SNR properties such as its morphology, shock velocity, and the ambient medium properties. The outline of the paper is as follows. We describe the observations and data reduction in Section \ref{obs}. Our analysis and results are reported in Section \ref{analysis}. Finally, we discuss our findings and summarize our main conclusions in Section \ref{discuss}.

\section{Observations and data reduction}
\label{obs}
\subsection{Imaging}

The photometric observations of the SNR G107.5$-$1.5 were performed with the Ritchey-Chrétien RTT150 telescope of T\"{U}B\.{I}TAK National Observatory (TUG)\footnote{\url{https://tug.tubitak.gov.tr}}, Turkey (36${^o}$ 49$\arcmin$ 30$\arcsec$ N, 02$^{\rm h}$ 01$^{\rm m}$ 20$^{\rm s}$ E, altitude 2500 m) between 2020 and 2021. CCD camera used in imaging consists of 2048 $\times$ 2048 pixels, each of 15 ${\mu}$m $\times$ 15 ${\mu}$m, covering 13$\times$13 arcmin$^2$ field of view (FoV).

In addition to the RTT150 telescope, the 1.0 m fully automatic Ritchey-Chrétien T100 telescope was also used for photometric observations of G107.5$-$1.5. The telescope has a large format CCD camera dedicated to wide-field imaging. This camera with 21.5$\times$21.5 arcmin$^2$ FoV has 4096$\times$4096 pixels, each of 15 ${\mu}$m $\times$ 15 ${\mu}$m.

We observed the radio shell, including the inner and outer regions of the SNR G107.5$-$1.5. However, we detected optical emission in only 4 regions: its North East (NE), North West (NW), South East (SE), and East (E) regions. In the observations made with both telescopes, H$\alpha$ and continuum filters have been used. Characteristics of the filters and details of the photometric observations of the observed regions are given in Table \ref{Table1}.

The raw data were processed by using standard \texttt{Image Reduction Analysis Facility} (\texttt{IRAF}) routines such as bias and dark frame subtraction, flat-field division, and bad-pixel correction.

\begin{table}
 \caption{Log of photometric observations and characteristics of the filters used in our observations. The exposure time is for a single frame. The numbers in brackets show the number of individual frames.}
 \begin{tabular}{@{}p{1.0cm}p{2.3cm}p{1.6cm}p{1.7cm}@{}}
 \hline
Region & Image centre  & Exposure & Observation  \\
 & ($\alpha; \delta$) &  time & date \\
  & (h m s; ${^o}$ $\arcmin$ $\arcsec$) &  (s) & (yyyy-mm-dd) \\
\hline
NW & 22 49 28; 57 49 48  & 900(5) & 2020-07-19  \\
NE & 22 54 31; 58 15 19  & 900(5) & 2020-11-16  \\
SE & 22 56 15; 57 46 04  & 600(1) & 2021-07-13  \\
E  &   22:55:33; 58:03:10 &  400(1) &  2022-08-31 \\
 \hline
 \hline
Telescope &Filter       			& Wavelength    	 & FWHM             \\ 
   &          			&    (nm)	             &   (nm)             \\ 
 \hline
RTT150&H$\alpha$     		    &     656.3           &   5                         \\
&H$\alpha$-cont     	    	&     644.6           &   13                   \\
\hline
T100&H$\alpha$     		    &     656.0           &   10.8                \\
&Cont blue                   &     551.0           &   88                     \\ 
\hline
\label{Table1}
\end{tabular}
\end{table}

\subsection{Spectroscopy}

Spectroscopic observations were carried out with the Faint Object Spectrograph and Camera (TFOSC) installed at the f/7.7 Cassegrain focus of the RTT150. In these observations, the grism number 15, which is most suitable for the spectral lines seen in SNRs, was chosen. The resolution and wavelength range for the selected grism are 749 and 3230 {\AA}–9120 {\AA}, respectively. Two spectroscopic techniques have been used whose details are given below. 

\subsubsection{Long-slit Spectroscopy}
Long-slit spectra were taken on the relatively bright optical emission at 7 different locations. The log of the spectroscopic observations is given in Table \ref{Table2}. The 134 $\mu$m slit was used for our observations. Each spectrum was reduced using \texttt {IRAF} according to the standard spectral reduction steps of bias subtraction, flat-field division, background-light subtraction, wavelength, and flux calibration. For wavelength calibration, Iron–Argon lamp spectra were taken after each object spectrum. Spectrophotometric standard stars  BD+75d325, Feige 34, and BD+28D4211  \citep{Ok90} were observed for flux calibration. After all these data processing steps, the effects of city lights and atmospheric molecular lines were extracted from the spectra.

\begin{table}
\centering
 \caption{Log of spectroscopic observations. The exposure time is for a single frame. The numbers in brackets show the number of individual frames.}
 \begin{tabular}{@{}p{0.6cm}p{2.8cm}p{1.6cm}p{1.8cm}@{}}
 \hline
Slit & Slit centre  & Exposure & Observation  \\
 & ($\alpha; \delta$) & time & date \\
  & (h m s; ${^o}$ $\arcmin$ $\arcsec$) &  (s) & (yyyy-mm-dd)\\
\hline
&\multicolumn{2}{c}{NW region}\\
\hline
1 & 22 49 38; 57 47 01  & 900(2) & 2020-07-11  \\
2 & 22 49 22; 57 45 58  & 900(1)  & 2020-07-22 \\
\hline
&\multicolumn{2}{c}{NE region}\\
\hline
1 & 22 53 41; 58 08 51  & 900(2) & 2020-08-16 \\
\hline
&\multicolumn{2}{c}{SE region}\\
\hline
1 & 22 56 10; 57 45 58  & 900(1)  & 2021-07-13 \\
2 & 22 56 12; 57 45 11  & 900(1) & 2021-07-18  \\
 \hline
 &\multicolumn{2}{c}{E region}\\
\hline
1 & 22:55:33; 58:02:29  & 900(3) & 2022-08-31  \\
2 & 22:55:27; 58:02:15  & 900(3) & 2022-08-31  \\
 \hline
\label{Table2}
\end{tabular}
\end{table}

\subsubsection{MOS Observations and Data Reduction}
NW region of G107.5$-$1.5 in which the optical emission was discovered has been observed with the Multi-Object Spectrograph (MOS)\footnote{\url{https://tug.tubitak.gov.tr/tr/icerik/tfosc-mos}} which runs as one of the observing modes of the TFOSC instrument attached to RTT150 telescope. During the MOS operation mode, grism-15\footnote{\url{https://tug.tubitak.gov.tr/sites/images/tug/gr15.gif}} providing an average resolving power of $R$ $\sim$ 750 in the wavelength range 3230$-$9120 \AA\,is used. Light from the telescope is fed to TFOSC through 100 $\mu$m pinholes on the mask, which correspond to the interesting points in the sky. Two masks (NW1 and NW2) each having 20 pinholes (19 for G107.5$-$1.5 and 1 for stellar source) were prepared before our observations. The stellar source is used for checking the wavelength calibration purpose. In our case, an $V\sim12$ mag star TYC-3992-57-1 is used as the stellar source. The MOS observations with the two masks have been performed on one observing night (2021-12-02). Each spectrum is a collection of five 900 s exposure, a total of 4500 s, which allowed us to detect and clean cosmic rays while combining.

Since the location of the pinholes on the mask is not the same, each spectrum has a unique wavelength coverage and dispersion solution. Therefore, besides the science spectrum, the Iron-Argon lamp was used as the wavelength calibration source with the mask on. Moreover, to see the variation of the dispersion all over the CCD field, slit spectra covering the whole CCD image have been obtained.

The reduction of the MOS spectra has been performed with the MOS Reduction Software (\texttt{mrs}; \citealt{Bakis2021}). \texttt{mrs} is a collection of open-source computer programs designed to reduce the MOS spectra and available on the net\footnote{\url{https://github.com/vbakis/mrs}}. It basically locates the pinholes on the observing mask and extracts the spectra on each aperture. In \texttt{mrs}, there is a dedicated algorithm for wavelength calibration of each aperture by cross-correlating the apertures with the calibration aperture which is a spectrum pre-calibrated with the slit spectrum of the calibration lamp source. The code, written in {\sc python}, provides a graphical user interface for an easy-to-use data reduction environment.

\section{Analysis and Results}
\label{analysis}
\subsection{Images}
\begin{figure}
\includegraphics[angle=0, width=9cm]{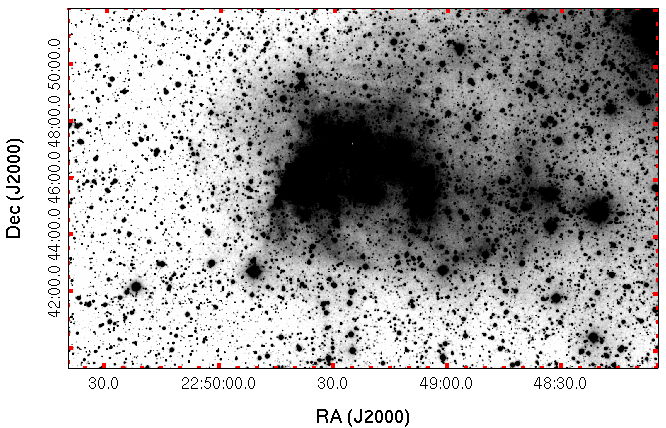}
\includegraphics[angle=0, width=9cm]{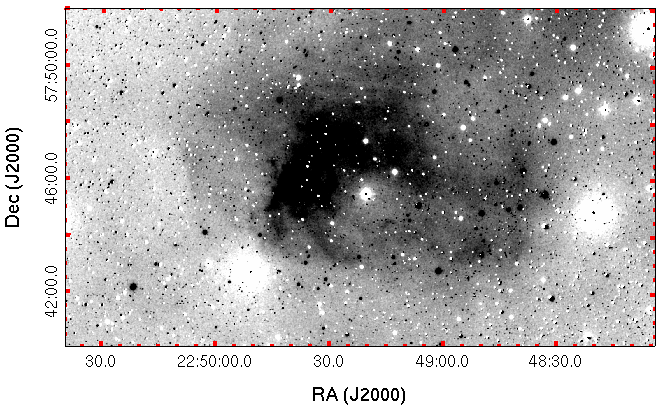}
\caption{The H$\alpha$ (upper panel) and continuum-subtracted H$\alpha$ image (lower panel) of the G107.5$-$1.5's NW region taken with the T100 telescope. In this figure, North is up, East is to the left.}
\label{figure1}
\end{figure}

\begin{figure}
\includegraphics[angle=0, width=9cm]{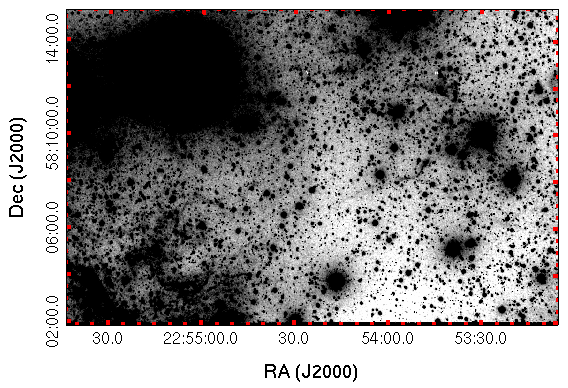}
\caption{The H$\alpha$ image of the G107.5$-$1.5's NE region taken with the T100 telescope. The directions are same as in Fig. \ref{figure1}.}
\label{figure2}
\end{figure}

\begin{figure*}
\centering
\includegraphics[angle=0, width=7cm]{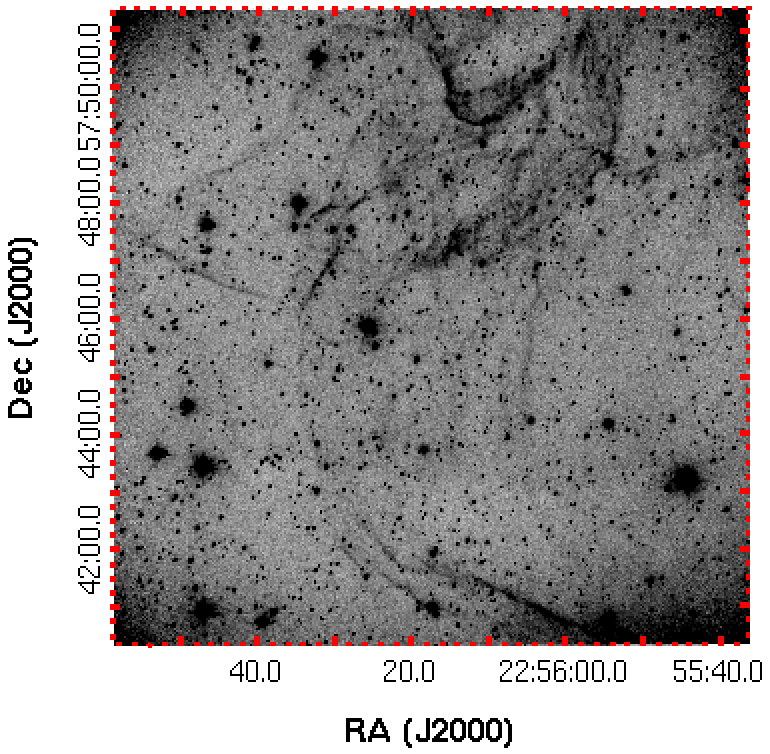}
\includegraphics[angle=0, width=7cm]{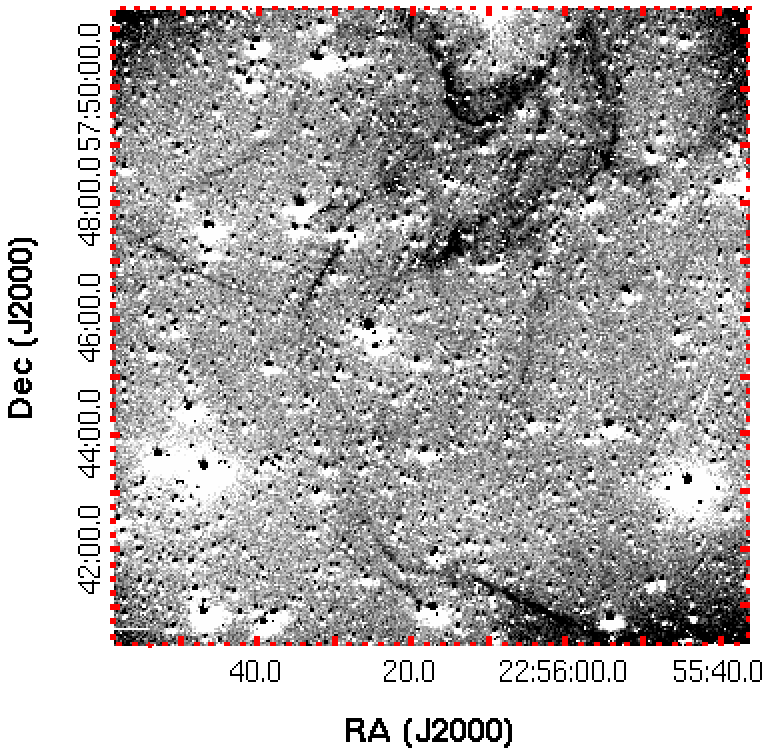}
\caption{The H$\alpha$ image (upper panel) and continuum-subtracted H$\alpha$ image (lower panel) of the G107.5$-$1.5's SE region taken with the RTT150 telescope. The directions are same as in Fig. \ref{figure1}.}
\label{figure3}
\end{figure*}

\begin{figure*}
\centering
\includegraphics[angle=0, width=9cm]{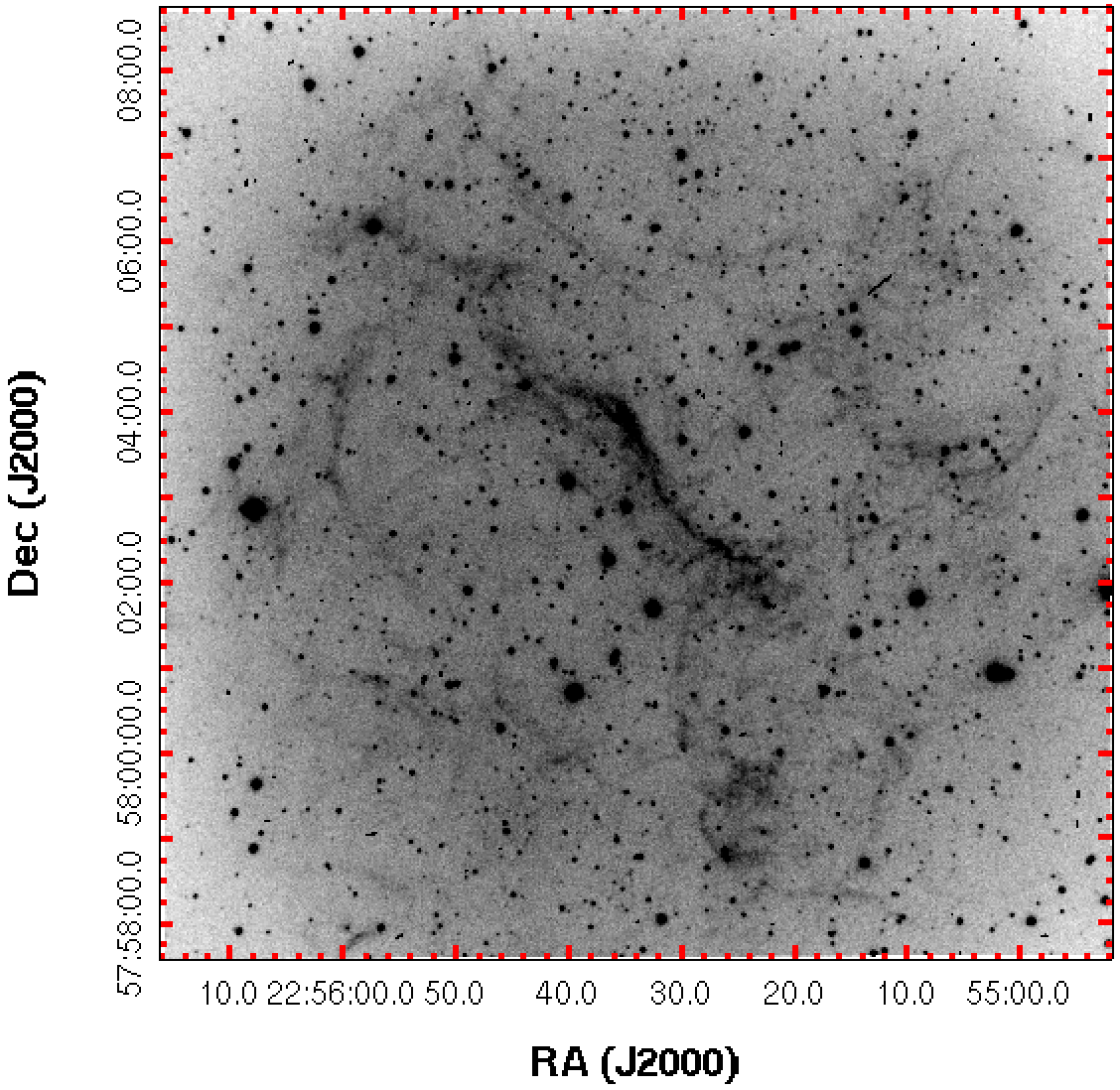}
\caption{The H$\alpha$ image of the E region taken with the RTT150 telescope. The directions are same as in Fig. \ref{figure1}}.
\label{figure3_E}
\end{figure*}

\begin{figure*}
\includegraphics[angle=0, width=15cm]{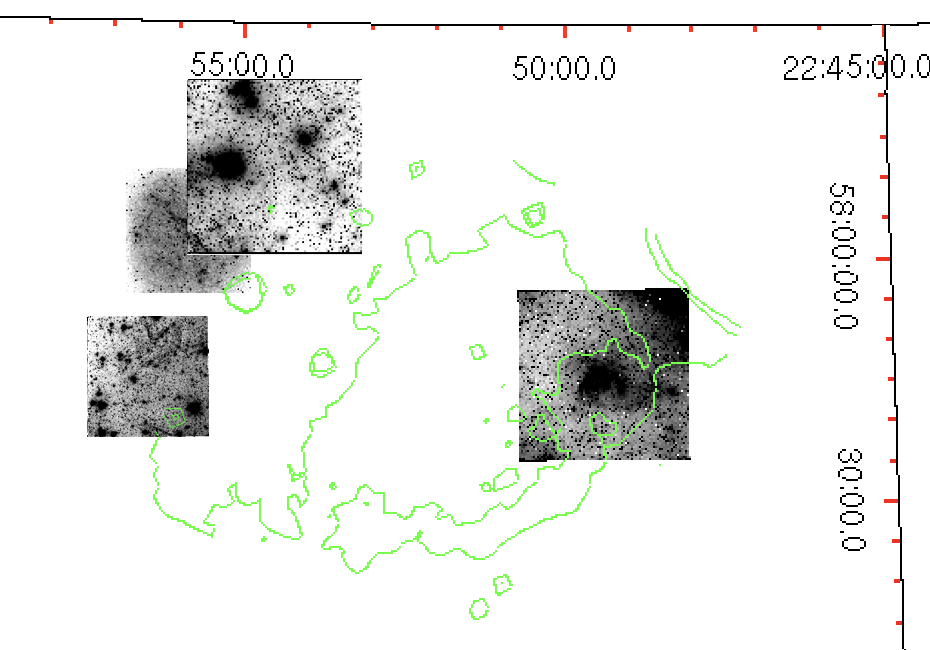}
\caption{The mosaic image combines optical images (H$\alpha$) of four regions (NW, NE, SE, and E) of the G107.5$-$1.5 overlaid with the radio-continuum contours (green) at 1420 MHz taken from the CGPS \citep{Ta03}. The radio-continuum contour levels range from 4.11 to 8.48 mJy beam$^{-1}$. The coverage of the optical images of the NE and NW areas are 21.5$\times$21.5 arcmin$^2$, while the SE and E areas are 11.1$\times$11.1 arcmin$^2$. In all figures, the directions are same as in Fig. \ref{figure1}, and the coordinates refer to J2000.0 epoch.}
\label{figure4}
\end{figure*}

We present the H$\alpha$ and continuum-subtracted H$\alpha$ images of NW and SE regions, and H$\alpha$ images of NE and E regions in Figs. \ref{figure1} $-$ \ref{figure3_E}. As can be seen in these images, the emission consists of filaments and a mostly diffuse structure. In Fig. \ref{figure4}, we compared the H$\alpha$ mosaic image of the observed regions with the radio continuum contours at 1420 MHz from the CGPS data \citep{Ta03}.

\subsection{Spectra}
\begin{figure*}
\includegraphics[angle=0, width=16cm]{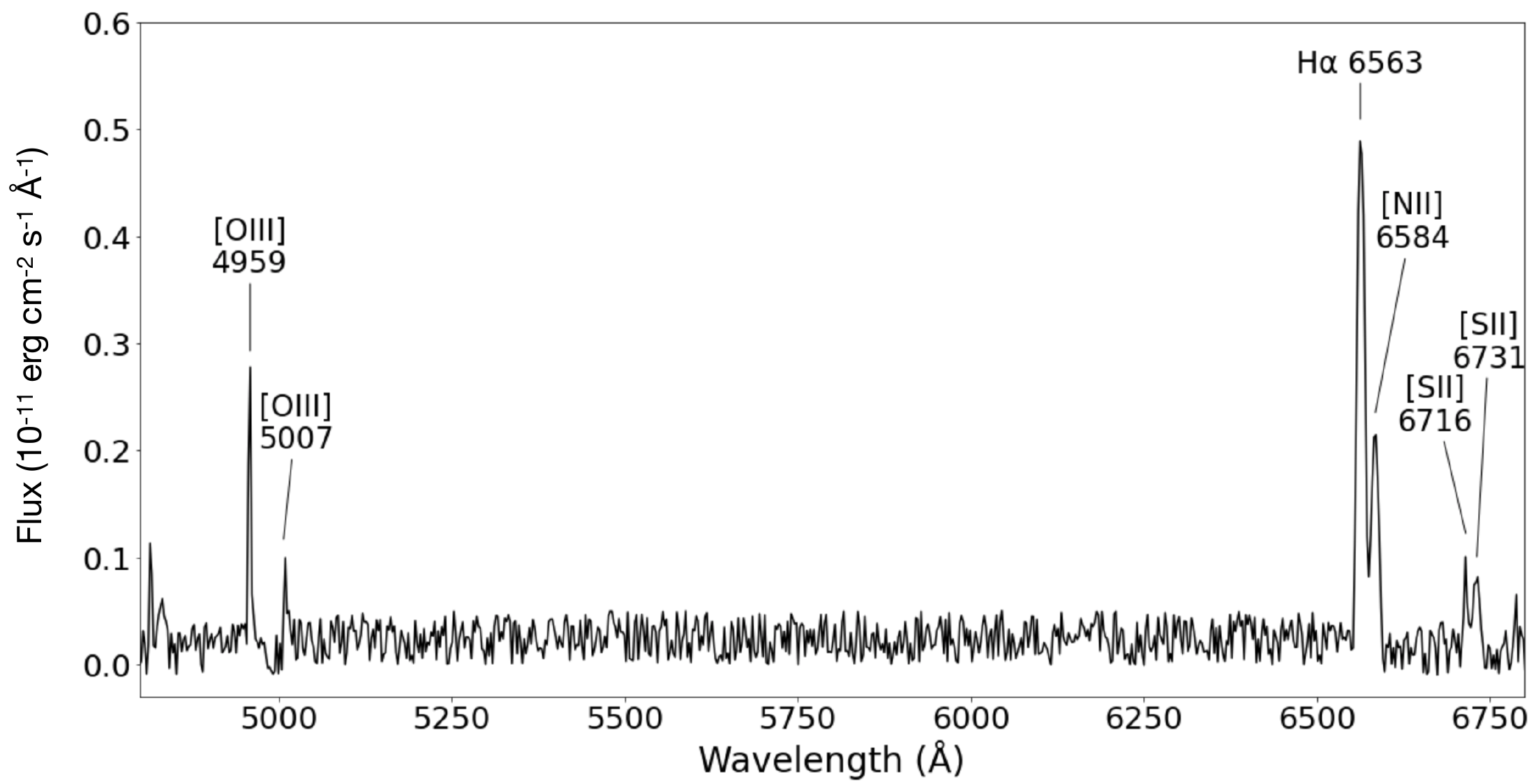}
\includegraphics[angle=0, width=16cm]{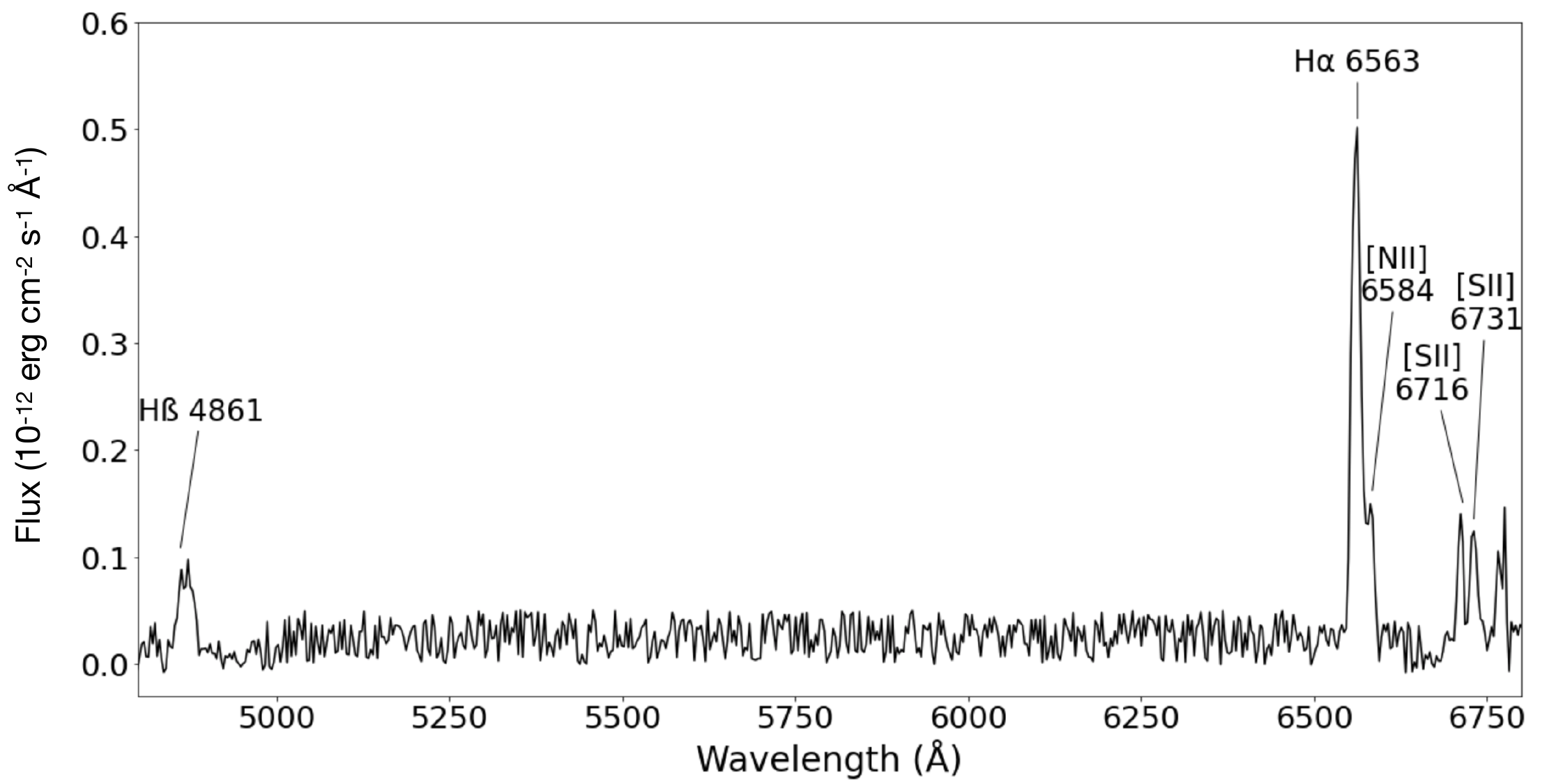}
\caption{Long-slit spectra extracted for 2 slit positions located in the NW region. The slit centres and exposure times are given in Table \ref{Table2}. The H$\beta$$\lambda$4861, [O\,{\sc iii}]$\lambda$4959, $\lambda$5007, H$\alpha$$\lambda$6563, [N\,{\sc ii}]$\lambda$6584 and [S\,{\sc ii}]$\lambda$6716, $\lambda$6731 lines are clearly visible.}
\label{figure5}
\end{figure*}

\begin{figure*}
\includegraphics[angle=0, width=16cm]{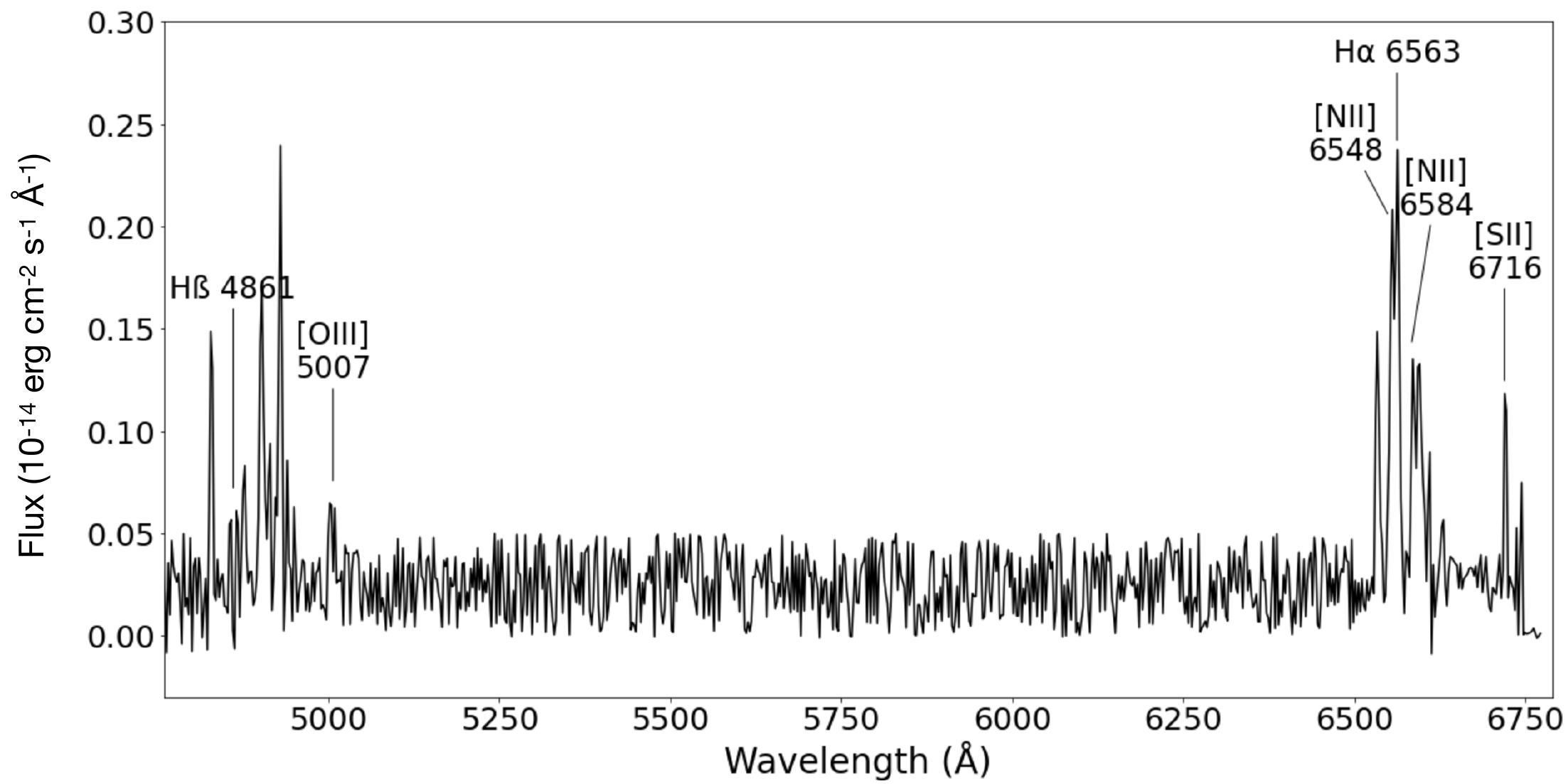}
\caption{Long-slit spectra extracted for a slit position located in the NE region. The slit centre and exposure time are given in Table \ref{Table2}. The H$\beta$$\lambda$4861, [O\,{\sc iii}]$\lambda$5007, H$\alpha$$\lambda$6563, [N\,{\sc ii}]$\lambda$6548, $\lambda$6584 and [S\,{\sc ii}]$\lambda$6716, $\lambda$6731 lines are clearly visible.}
\label{figure6}
\end{figure*}

\begin{figure*}
\includegraphics[angle=0, width=16cm]{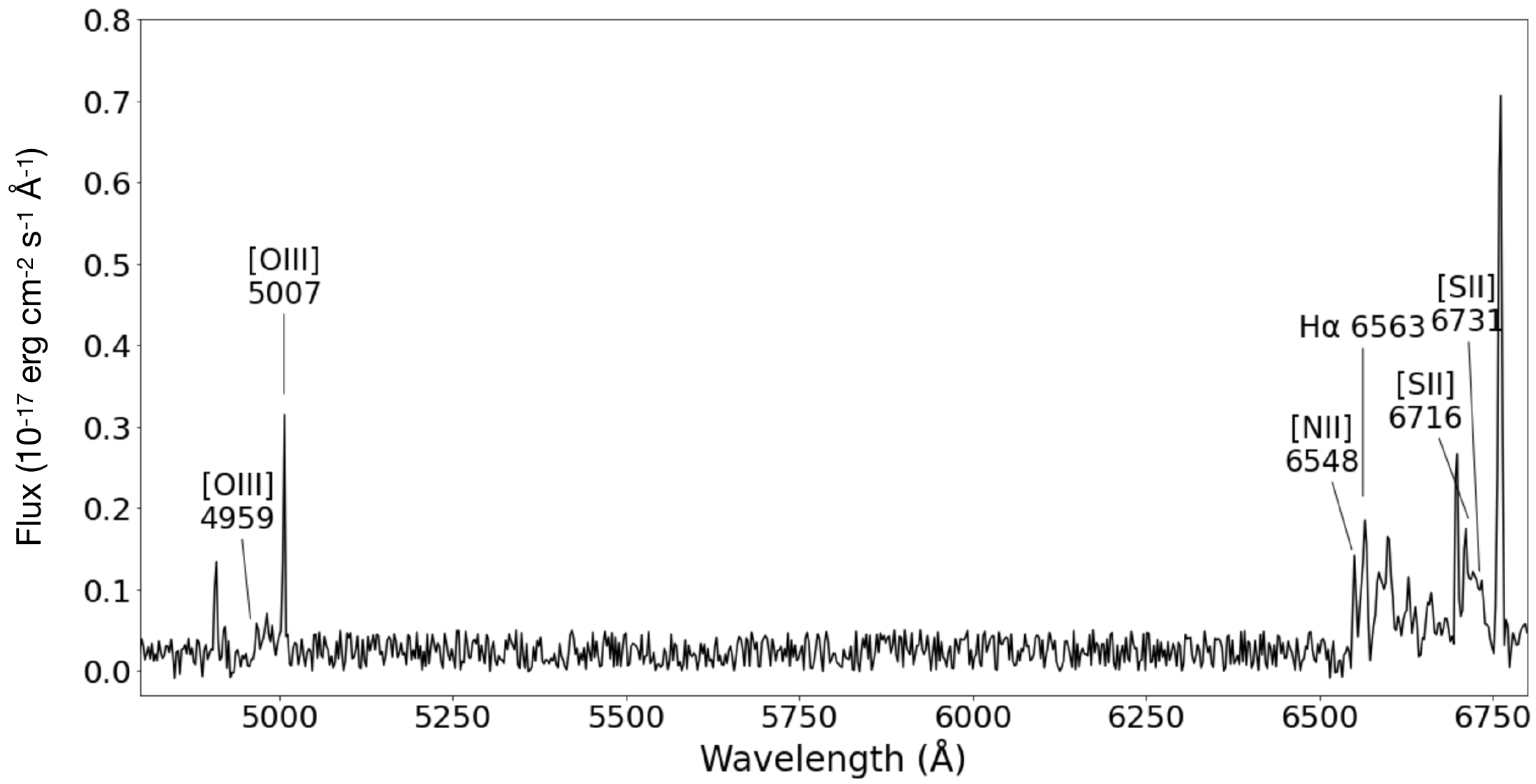}
\includegraphics[angle=0, width=16cm]{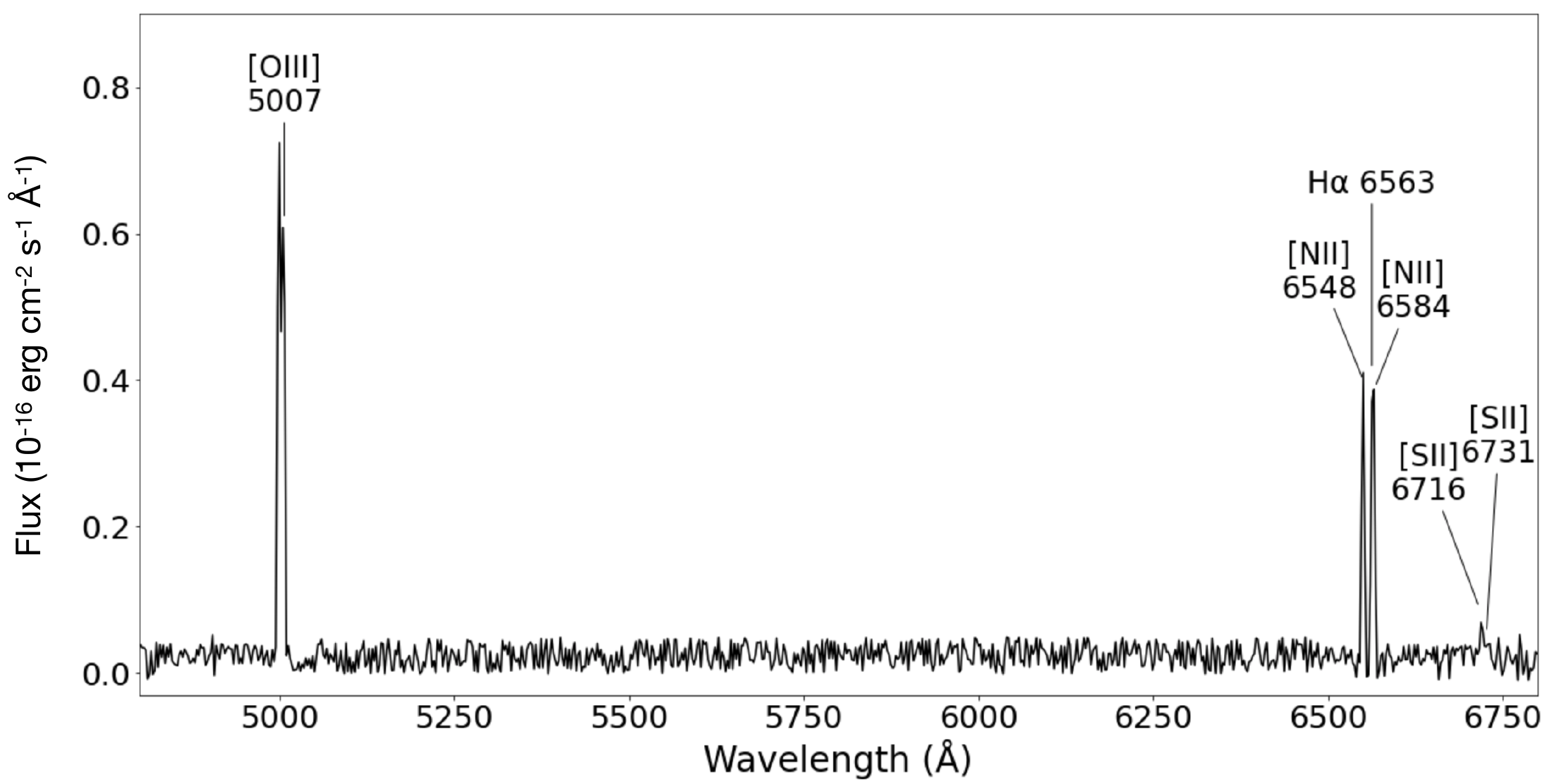}
\caption{Long-slit spectra extracted for 2 slit positions located in the SE region. The slit centres and exposure times are given in Table \ref{Table2}. The [O\,{\sc iii}]$\lambda$4959, $\lambda$5007, H$\alpha$$\lambda$6563, [N\,{\sc ii}]$\lambda$6584, $\lambda$6584 and [S\,{\sc ii}]$\lambda$6716, $\lambda$6731 lines are clearly visible.}
\label{figure7}
\end{figure*}

\begin{figure*}
\includegraphics[angle=0, width=16cm]{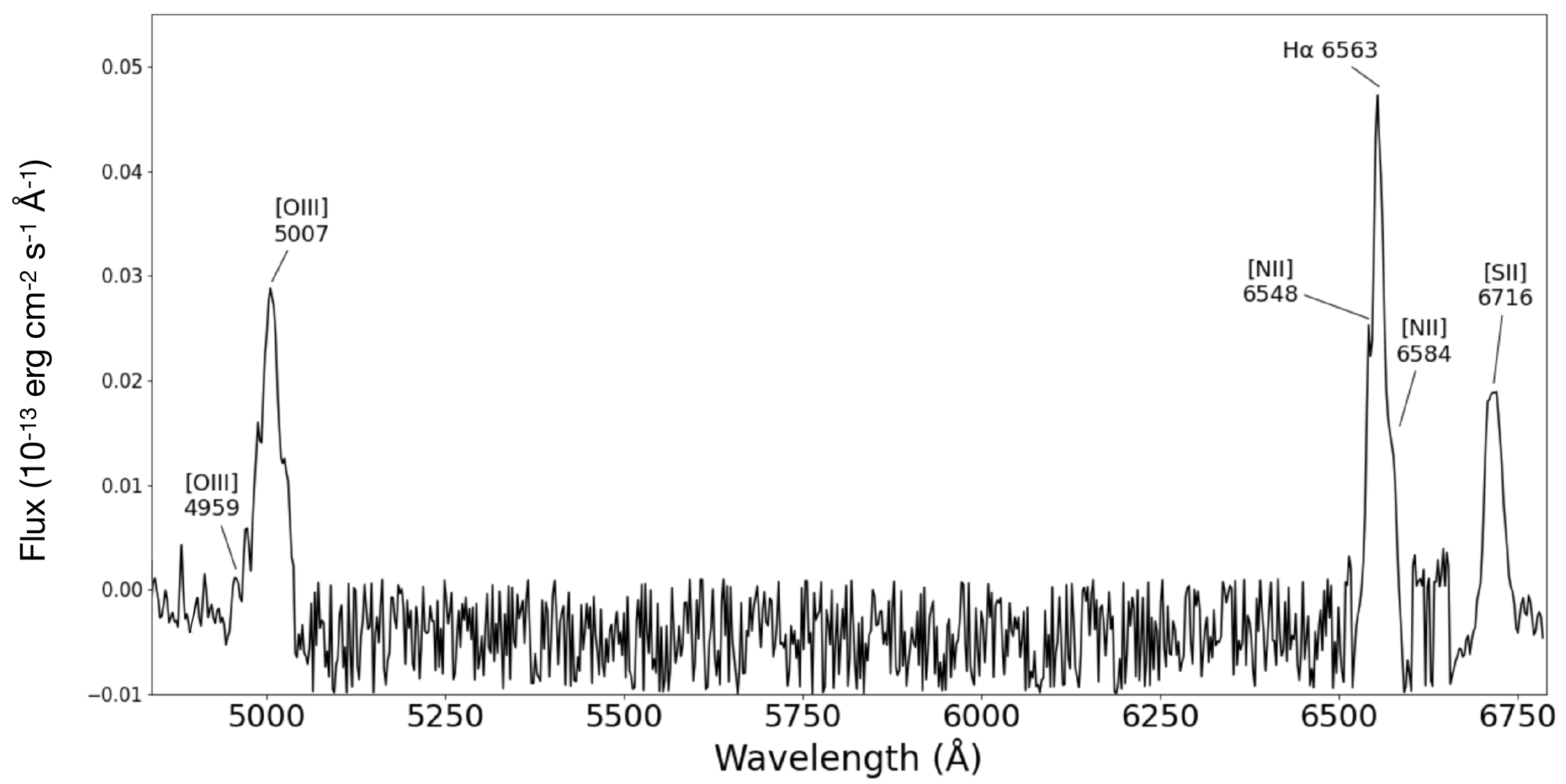}
\includegraphics[angle=0, width=16cm]{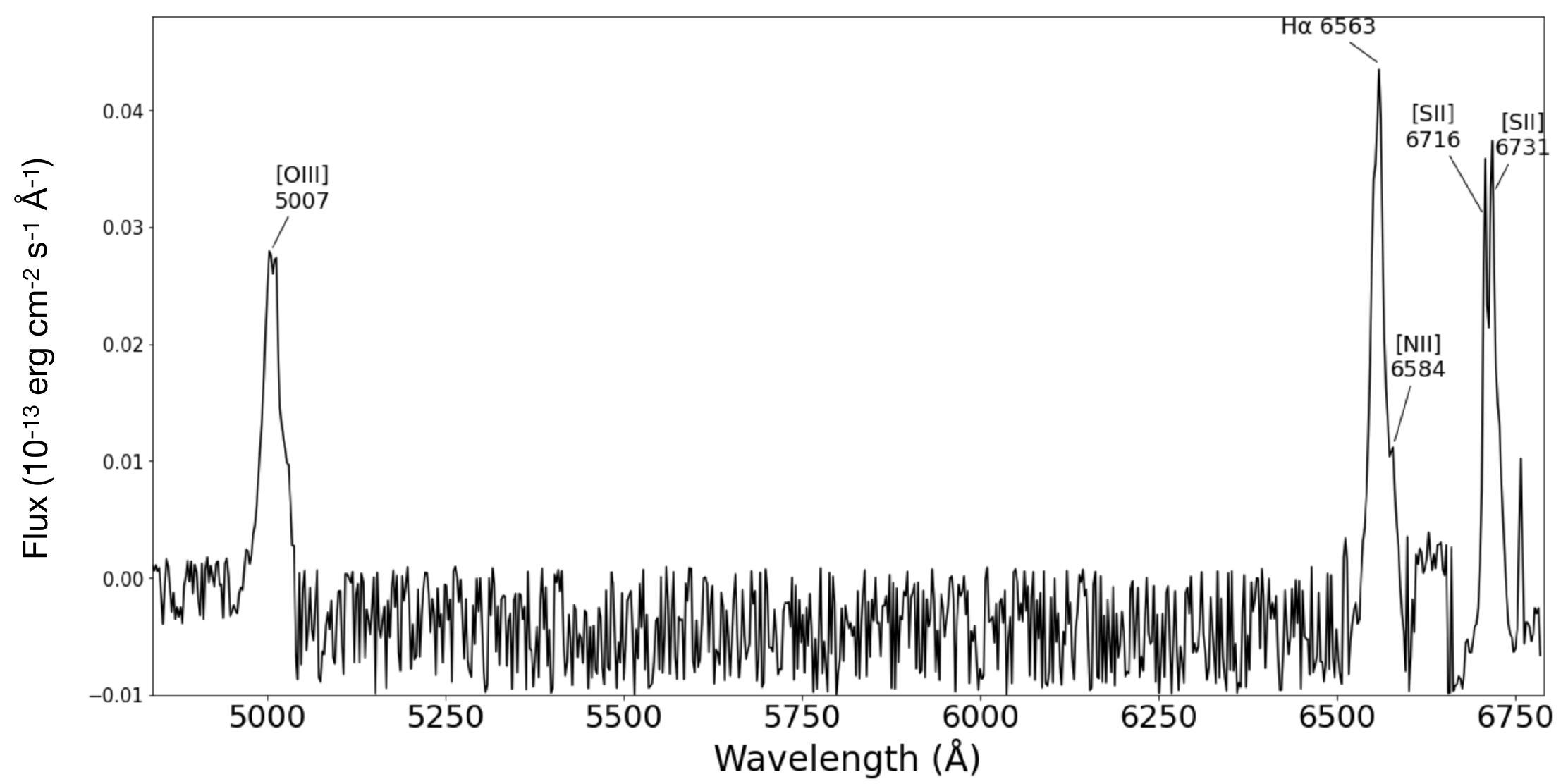}
\caption{
Long-slit spectra extracted for 2 slit positions located in the E region. The slit centres and exposure times are given in Table \ref{Table2}. The [O\,{\sc iii}]$\lambda$4959, $\lambda$5007, H$\alpha$$\lambda$6563, [N\,{\sc ii}]$\lambda$6584, $\lambda$6584 and [S\,{\sc ii}]$\lambda$6716, $\lambda$6731 lines are clearly visible.}
\label{figure_devam}
\end{figure*}

Our long-slit spectra for NW, NE, SE and E regions are given in Figs. \ref{figure5} $-$ \ref{figure_devam}, respectively. For these regions, the fluxes of lines, line ratios, and some physical parameters are presented in Table \ref{Table3}.

\begin{table*}
\centering
 \caption{Observed line fluxes ($F$) for NW, NE, SE and E regions of G107.5$-$1.5 were obtained with the long-slit spectra. All fluxes are normalised $F$(H$\alpha$) = 100. The signal-to-noise ratios (S/N) are given between brackets. The line ratios and physical parameters are given at the bottom of the table.}
 \label{Table3}
 \begin{threeparttable}
 \begin{tabular}{@{}lccccccc@{}}
 \hline
Lines (\AA) & \multicolumn{7}{c}{Fluxes (H$\alpha$=100) and S/N ratios} \\
\hline
& &\multicolumn{1}{l}{NW} & \multicolumn{1}{c}{NE} & \multicolumn{2}{c}{SE} & \multicolumn{2}{c}{E}   \\
     \cline{2-8}   
&  1  & 2  & 1 &  1  & 2 &  1  & 2 \\
\hline
H$\beta$ ($\lambda$4861)& $-$  & 11 (15)  & 13  (30)  &  $-$ & $-$ &  $-$ & $-$  \\

$[$O$\,${\sc iii}$]$ ($\lambda$4959) & 20 (47) & $-$ & $-$  &  13 (17) &$-$   &  12 (5)&$-$   \\

$[$O$\,${\sc iii}$]$ ($\lambda$5007) & 5 (17) & $-$  &  7 (34)   &  59 (28) & 136 (30)& 81 (15) &104 (18)  \\

$[$N$\,${\sc ii}$]$ ($\lambda$6548) & $-$ & $-$   & 113  (29)  & 39 (41)  & 87 (50)& 28(13) &58(15)  \\

H$\alpha (\lambda$6563) & 100 (40)  & 100 (39)   & 100 (34)   & 100 (50) & 100 (47)& 100(24) &100(22) \\

$[$N$\,${\sc ii}$]$ ($\lambda$6584) & 39 (18) & 24 (51)   & 23  (36)  &  $-$  & 44 (10) &32 (7) &27(5) \\

$[$S$\,${\sc ii}$]$ ($\lambda$6716) & 9 (10) & 17 (28)     & 27 (24) & 75 (38) & 17 (18)&77 (10)&41 (20)\\

$[$S$\,${\sc ii}$]$ ($\lambda$6731) & 14 (26) & 21 (12)     & $-$  &  60 (20) & 13 (10) &$-$& 33(20)\\
\\[0.5 ex]
\hline
Line Ratios &\multicolumn{7}{c}{Values}\\
\hline
\\[0.5 ex]
F (H$\alpha$) (erg cm$^{-2}$ s$^{-1}$) &   6.42$\times$$10^{-11}$ & 7.9$\times$$10^{-12}$ & 1.98$\times$$10^{-14}$ &  0.18$\times$$10^{-17}$ & 2.41$\times$$10^{-16}$ &  1.2$\times$$10^{-13}$ & 1.3$\times$$10^{-13}$ \\

[S\,{\sc ii}] / H$\alpha$ &   0.23 $\pm$ 0.03 & 0.38 $\pm$ 0.02  &  0.27 $\pm$ 0.05  & 1.35 $\pm$ 0.03  & 0.30 $\pm$ 0.04 & 0.77 $\pm$ 0.02& 0.75 $\pm$ 0.02\\

 [N\,{\sc ii}] / H$\alpha$   & 0.39 $\pm$ 0.02    & 0.24 $\pm$ 0.01  & 1.36 $\pm$ 0.02   & 0.39 $\pm$ 0.15 &  1.31 $\pm$ 0.16  &0.60 $\pm$ 0.06 &0.85$\pm$ 0.04  \\

[S\,{\sc ii}]$\lambda6716/\lambda$6731 &   0.64 $\pm$ 0.04 & 0.81 $\pm$ 0.06  & $-$  & 1.24 $\pm$ 0.04 & 1.25 $\pm$ 0.09  &$-$&1.15 $\pm$ 0.04 \\

[O\,{\sc iii}] ($\lambda$4959$ + \lambda$5007)/H$\beta$ $\lambda$4861 &   $-$  & $-$  &  0.54 $\pm$ 0.02  & $-$ & $-$ &$-$   &$-$ \\
\\[0.5 ex]
\hline
Parameters &\multicolumn{7}{c}{Values}\\
\hline
\\[0.5 ex]

$N_{\rm e}$(cm$^{-3}$)&   3301$\pm$ 530  & 1432$\pm$ 150  & $-$  & 240$\pm$3 & 231$\pm$6  &$-$& 262$\pm$7 \\

$V_{\rm s}$ (km s$^{-1}$) &   $-$  & $-$  &   80  & $-$ & $-$  &$-$  &$-$  \\

$E(B-V)$$^{\dagger}$&   $-$  & 0.94 $\pm$ 0.02  & 0.81 $\pm$ 0.04  & $-$ & $-$   \\

$A_{\rm v}$ &   $-$ & 2.91 $\pm$ 0.05  & 2.50$\pm$ 0.11   & $-$ & $-$ &$-$  &$-$  \\

$N_{\rm H}$ ($10^{21}$ cm$^{-2}$) &   $-$  & 5.07$\pm$ 0.08  & 4.38$\pm$ 0.19  &  $-$ & $-$ &$-$  &$-$   \\
\hline
\end{tabular}
\begin{tablenotes}
\small
\item {$^{\dagger}$ $E(B-V)=0.664c$, where $c$ is a logarithmic extinction determined as 1/0.331 log[(H$\alpha$/H$\beta$)/3] \citep{Ka76, Al84}.} \\
\end{tablenotes}
\end{threeparttable}
\end{table*}

\begin{figure*}
\centering
\begin{subfigure}[b]{\textwidth}
    \centering
    \includegraphics[width=0.7\textwidth]{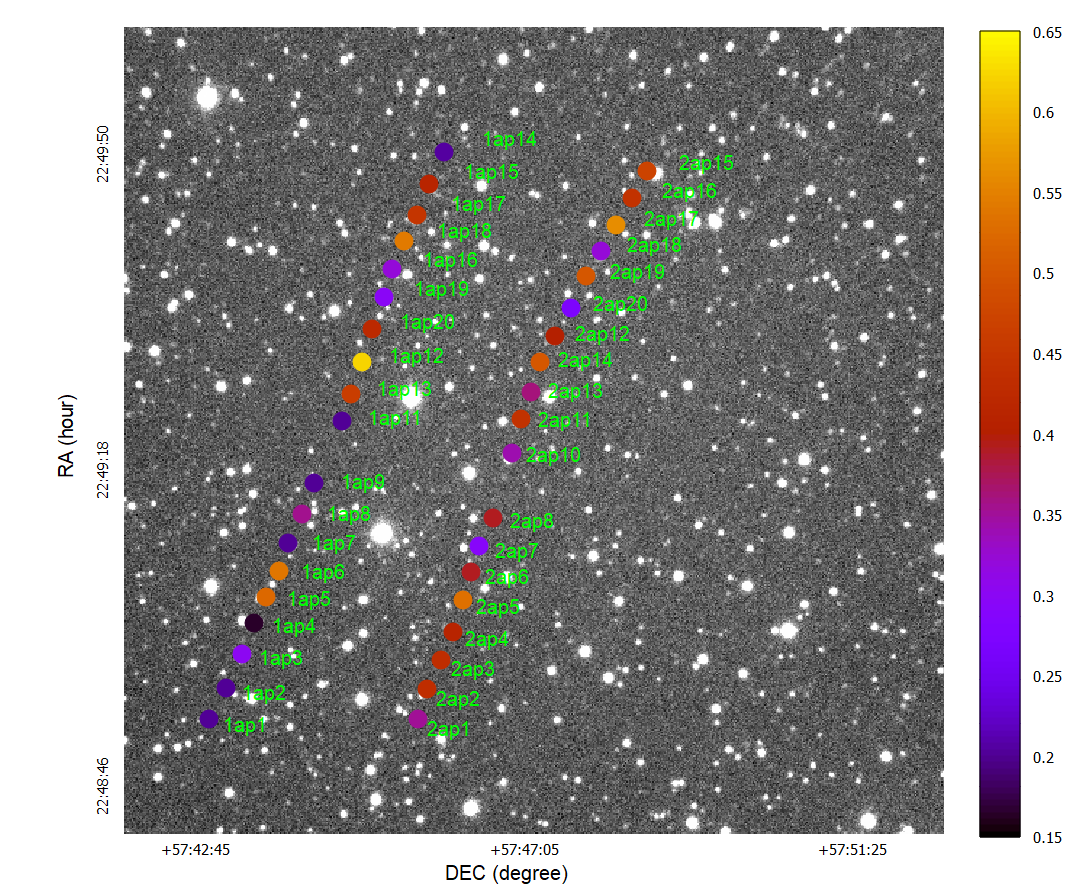}
\end{subfigure}
\newline
\begin{subfigure}[b]{0.3\textwidth}
    \centering
    \includegraphics[width=\textwidth]{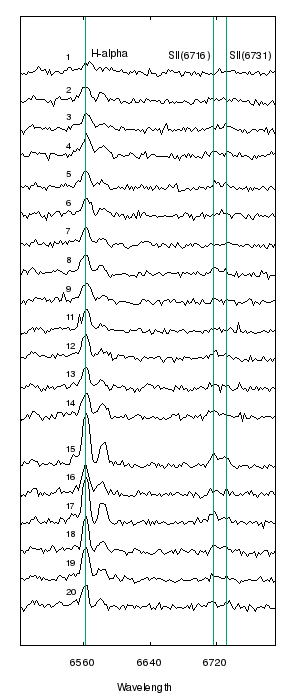}
\end{subfigure}
%\hfill
\begin{subfigure}[b]{0.3\textwidth}
    \centering
    \includegraphics[width=\textwidth]{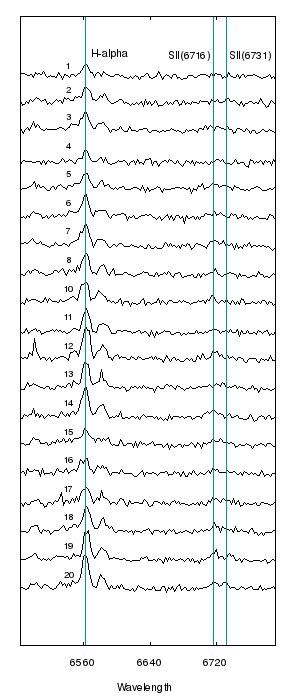}
\end{subfigure}
\caption{\textit{Top:} MOS apertures are marked on the field. The color bar indicates the magnitude of [S\,{\sc ii}]/H$\alpha$ ratio. \textit{Bottom:} H$\alpha$ and [S\,{\sc ii}] line profiles with aperture numbers indicated. The fields NW1 and NW2 are shown on each panel.}
\label{figure8}
\end{figure*}

The locations of the pinholes from two masks for the field of G107.5$-$1.5 are shown in the top panel of Fig. \ref{figure8}, where the background sky image is taken by RTT150 telescope without the filter. In the bottom panel of Fig. \ref{figure8}, we present the MOS spectra in the 6485$-$6790 {\AA} wavelength range, showing the H$\alpha$ $\lambda$6563{\AA}, and [S\,{\sc ii}] $\lambda \lambda$6716/6731{\AA} lines. The {[S\,{\sc ii}] / H$\alpha$} line ratios for the NW1 and NW2 regions of G107.5$-$1.5 obtained with the MOS spectra are given in Table \ref{Table4}.

\begin{table}
 \caption{{[S\,{\sc ii}] / H$\alpha$} line ratios for NW region of G107.5$-$1.5 obtained with the MOS spectra.}
 \label{Table4}
 \begin{tabular}{@{}p{2.0cm}p{2.5cm}p{2.5cm}@{}}
\\[0.05 ex]
\hline
    Hole  &\multicolumn{2}{c}{[S\,{\sc ii}] / H$\alpha$}\\
     \cline{2-3}
 &  MASK 1   & MASK 2  \\
\hline
1 &   0.36 $\pm$ 0.01 &  0.35	$\pm$	0.02 \\

2 &     0.48 $\pm$ 0.05 &   0.43	$\pm$	0.01 \\

3 &   0.30 $\pm$ 0.05 &     0.42	$\pm$	0.03\\

4 &    0.17 $\pm$ 0.01  &   0.35	$\pm$	0.01\\

5 &   0.53 $\pm$ 0.04 & 0.54	$\pm$	0.02 \\

6 &    0.54 $\pm$ 0.03 & 0.38	$\pm$	0.01\\

7 &  0.32 $\pm$ 0.01    &   0.18	$\pm$	0.01\\

8 &   0.34 $\pm$ 0.01  &    0.39     $\pm$0.04\\

9 &   0.26 $\pm$ 0.03 & 0.31	$\pm$	0.02 \\

11 &   0.13 $\pm$ 0.01 & 0.44	$\pm$	0.02 \\

12 &   0.62 $\pm$ 0.01 & 0.39	$\pm$	0.01 \\

13 &  0.44 $\pm$ 0.04 & 0.35	$\pm$	0.02\\

14 &   0.21 $\pm$ 0.01 &    0.49     $\pm$ 0.04 \\

15 &   0.41 $\pm$ 0.02 &    0.48	$\pm$	0.02 \\

16 &   0.32 $\pm$ 0.05 &     0.44	$\pm$	0.04 \\

17 &   0.41 $\pm$ 0.02 & 0.56	$\pm$	0.02 \\

18 &   0.53 $\pm$ 0.01 & 0.30	$\pm$	0.03 \\

19 &  0.29 $\pm$ 0.02 & 0.50	$\pm$	0.03 \\

20 &   0.42 $\pm$ 0.02 & 0.27	$\pm$	0.01 \\
\hline
\hline
\end{tabular}
\begin{tablenotes}
\item 
\end{tablenotes}
\end{table}

\section{Discussion}
\label{discuss}
 
\subsection{Optical morphology of G107.5$-$1.5}

In order to investigate the optical morphological structure of G107.5$-$1.5, we performed narrow-band H$\alpha$ imaging of the SNR, including the inner and outer parts of the radio shell. We detected optical emission in the NW, NE, SE and E regions. As can be seen in Figs. \ref{figure1} and \ref{figure2}, the H$\alpha$ image of the NW and NE regions exhibits a diffuse emission structure. The H$\alpha$ emission of the SE and E regions shows long, curved, and parallel filaments, which shows a classical SNR morphological structure (see Figs. \ref{figure3} and \ref{figure3_E}). As can be seen in Fig. \ref{figure4}, the radio shell mostly overlaps the optical emission of the NW region, while the NE, SE and E regions are outside of the radio shell. The location and orientation of the optical filaments coincide with the radio dark parts.

\subsection{Optical spectroscopy of G107.5$-$1.5}
%General
Our optical long-slit spectra show hydrogen Balmer lines, H$\beta$ $\lambda$4861{\AA}, H$\alpha$ $\lambda$6563{\AA}, and forbidden lines [O\,{\sc iii}] $\lambda \lambda$4959/5007{\AA}, [N\,{\sc ii}] $\lambda \lambda$6548/6584{\AA}, and [S\,{\sc ii}] $\lambda \lambda$6716/6731{\AA}. 

%SII/Halpha ratio
For the SE and E regions, the ratio of [S\,{\sc ii}]/H$\alpha$ $>$ 0.5 supports the origin of the emission being from shock-heated gas \citep{Fe85}. For the NW and NE regions, [S\,{\sc ii}]/H$\alpha$ ratios lower than 0.5 from long-slit spectra are consistent with emission from the H\,{\sc ii} region.

%NII/Halpha ratio
The [N\,{\sc ii}]/H$\alpha$ ratio could also be used together to best separate photoionization and shock-ionization mechanisms (e.g. \citealt{Fe85, Frew2010}). As can be seen in Table \ref{Table3}, the [N\,{\sc ii}]/H$\alpha$ ratios are found to be $>$0.5 for the NE, E and and SE (slit 2) regions, which suggests the origin of the emission from shock-heated gas.

% sulphur lines 6717/6731 ratio and electron density
The ratio of [S\,{\sc ii}] 6717/6731 allows us to determine the electron density ($N_{\rm e}$) at the observed position \citep{OsFe06}. We estimated the electron density based on [S\,{\sc ii}] 6717/6731 flux ratio using the \texttt{temden} task of the \texttt{nebular} package, assuming an electron temperature ($T$) of $10^{4}$ K. For the NW region, we estimated an average electron density of $\sim$2400 cm$^{-3}$, which can be attributed to the dense ionized gas. On the other hand, in the SE and E regions, the average density is $\sim$235 cm$^{-3}$ and $\sim$262 cm$^{-3}$, respectively.

% O lines and shock velocity
The [O\,{\sc iii}]/H$\beta$ ratio is mainly an indicator of the mean level of ionization and temperature \citep{Do84}. For the NE region, the observed ratio of [O\,{\sc iii}]/H$\beta$ $\sim$ 0.54 indicates a shock velocity of 80 km s$^{-1}$, and also suggests that shocks with a complete recombination zone (see \citealt{Ra79, Sh79}). The absence of the [O\,{\sc iii}] 5007{\AA} emission line in the NW region (slit 2) implies the presence of shock with even lower velocity ($<$80 km s$^{-1}$). The weak [O\,{\sc iii}] emission may be explained by slow shocks propagating into the interstellar medium (ISM) \citep{Ra88}.

%reddening
We derived a logarithmic extinction $c$ for the NW and NE regions from the observed H$\alpha$/H$\beta$ flux ratio. We estimated reddening $E(B-V)$ of $\sim$0.94 and $\sim$0.81 for NW and NE region, respectively (see Table \ref{Table3}). Using the relation $N_{\rm H}$ = 5.4 $\times$ $10^{21}$ $\times$ $E(B-V)$ \citep{Pr95}, we calculated the column density of $\sim$4.7$\times$ $10^{21}$ cm$^{-2}$ is consistent with the total galactic column density of $\sim$6$\times$ $10^{21}$ cm$^{-2}$ \citep{DiLo90}. This implies that the H\,{\sc ii} regions in the NW and NE regions are much farther away than the SNR, and optical emission from these regions is not related to G107.5$-$1.5.

To explain the peculiar shape of the SNR, \citet{Kothes03} examined images of H\,{\sc i} and CO from the CGPS database, and concluded that there is no molecular material in the ambient medium, also found that there is a depression in the H\,{\sc i} emission at the location of the radio shell. The author argued that the bright shell expanding in a moderately dense ambient medium, while the other parts of the remnant expands into a very dense environment and decelerated very quickly. We found a shock velocity of $V_{\rm s}$ $\sim$  80 km s$^{-1}$ and an electron density of $n_{\rm e}$ $\sim$ 200 cm$^{-3}$, indicating that the SNR is in a late stage of evolution and the shock wave is expanding in a high ambient density medium. This result is consistent with the result of \citet{Kothes03}. 

\begin{figure*}
\includegraphics[angle=0, width=16cm]{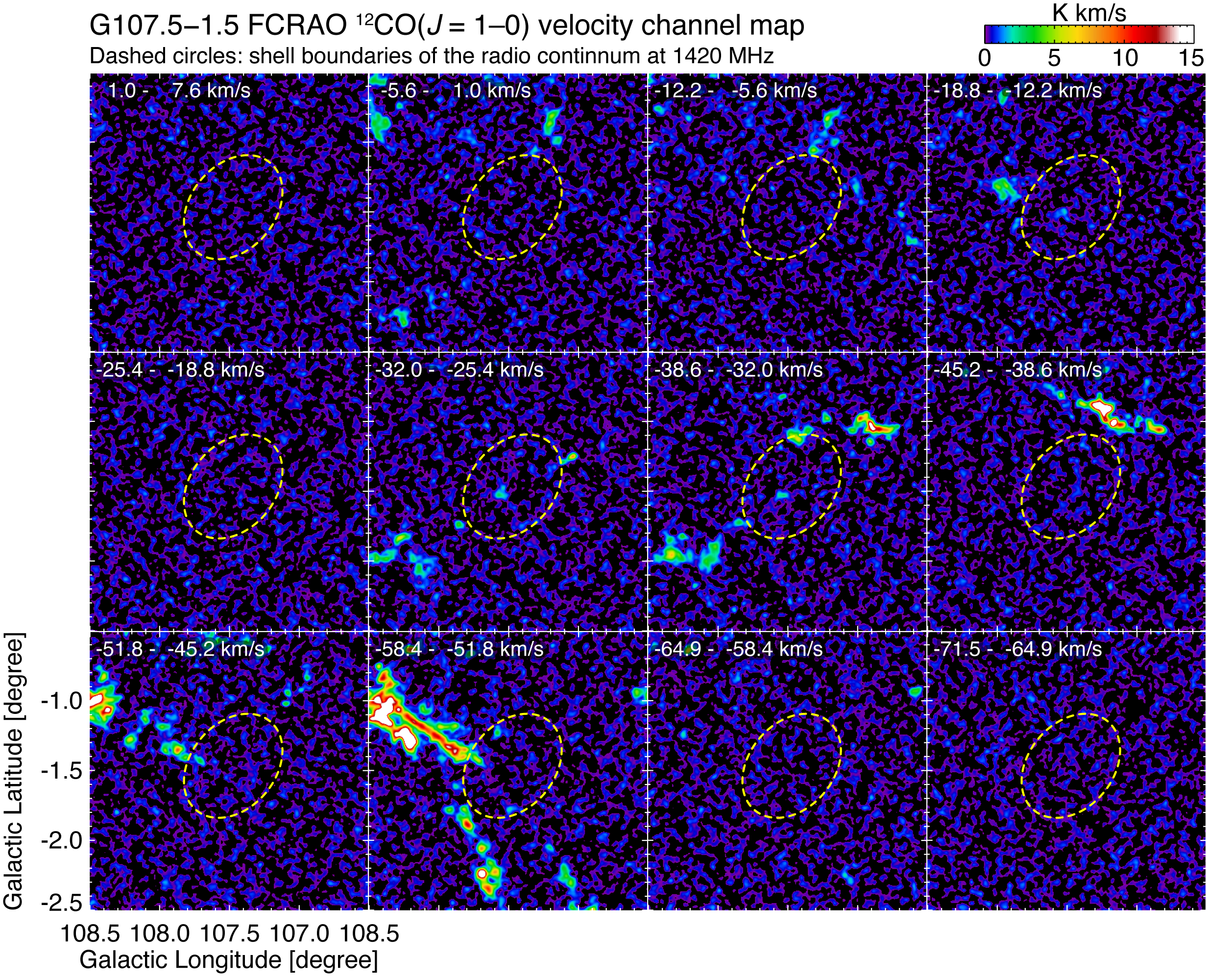}
\caption{Velocity channel maps of $^{12}$CO($J$~=~1--0) taken by FCRAO \citep{Ta03}. Each panel shows CO distribution every 6.6~km~s$^{-1}$ in the velocity range from $-64.9$ to 1.0~km~s$^{-1}$. The superposed dashed circles indicate the shell boundaries of the radio continuum at 1420~MHz \citep{Ta03}.}
\label{figure11}
\end{figure*}

We also confirmed that there are no dense molecular clouds toward G107.5$-$1.5 using the Five College Radio Astronomy Observatory (FCRAO) datasets \citep{Ta03}. Figure~\ref{figure11} shows the velocity channel maps CO toward the SNR. Although some clumpy clouds are spatially overlapped with the SNR shell boundary, no line-broadening of CO caused by the shock acceleration was observed.

In summary, we have presented the first results from an imaging and a spectroscopic survey of the optical emission associated with G107.5$-$1.5. The images taken with the H$\alpha$ filter reveal diffuse and filamentary structures. The filament-rich optical emission structure in the SE and E regions is morphologically consistent with a SNR nature. The spectra of the SE and E regions suggest that the optical emission originates from shock-heated gas ([S\,{\sc ii}]/H$\alpha$ $>$ 0.5). For the NW and NE regions, the average [S\,{\sc ii}]/H$\alpha$ ratio ([S\,{\sc ii}]/H$\alpha$ $\sim$ 0.3$-$0.4) indicates that the optical emission originates from ionized gas in H\,{\sc ii} regions. Our estimated electron densities for the NW region indicate that G107.5$-$1.5 interacts with a dense ionized gas. Further optical observations and in other passbands, especially in X-rays, could give more detailed information on the nature of this SNR and the interaction between the SNR and the surrounding ISM.

\section*{Acknowledgements}
We thank T\"{U}B\.{I}TAK National Observatory for partial support in using RTT150 (Russian-Turkish 1.5-m telescope in Antalya) with project number 1562 and T100 telescope with project number 1592. We also thank the anonymous referee for useful comments and suggestions that helped to improve the paper.

\section*{DATA AVAILABILITY}
The optical photometric and spectral data obtained at
TUG used in this study will be made available by the corresponding author upon request.

%%% FACILITIES
 
%%%%%%%%%%%%%%%%%%%% REFERENCES %%%%%%%%%%%%%%%%%%

% Don't change these lines
\bsp	% typesetting comment
\label{lastpage}

\newpage

\begin{thebibliography}{99}

\bibitem[\protect\citeauthoryear{Aller}{1984}]{Al84}Aller L. H., 1984, Physics of thermal gaseous nebulae (D. Reidel Publishing Company)

\bibitem[\protect\citeauthoryear{Bakı\c{s}}{2021}]{Bakis2021} Bakı\c{s} V., 2021, {\color {navy}TJAA, 2, 21}

\bibitem[\protect\citeauthoryear{Blair \& Long}{1997}]{BlLo97} Blair W.~P., Long K.~S., 1997, {\color {navy}ApJS, 108, 261 }

\bibitem[\protect\citeauthoryear{Boumis et al.}{2009}]{Bo09}Boumis P., Xilouris E. M., Alikakos J., Christopoulou P. E., Mavromatakis F., Katsiyannis A. C., Goudis C. D., 2009, {\color {navy} A\&A, 499, 789}

\bibitem[\protect\citeauthoryear{Condon et al.}{1994}]{Co94} Condon J. J., Broderick J. J., Seielstad G. A., Douglas K., Gregory P. C., 1994, {\color {navy} AJ, 107, 1829}


\bibitem[\protect\citeauthoryear{Dickey \& Lockman}{1990}]{DiLo90}Dickey J. M., Lockman F. J., 1990, {\color {navy}ARAA, 28, 215}

\bibitem[\protect\citeauthoryear{Dopita et al.}{1984}]{Do84} Dopita M.~A., Binette L., Dodorico S., Benvenuti P., 1984, {\color {navy}ApJ, 276, 653} 

\bibitem[\protect\citeauthoryear{Dopita et al.}{2010}]{Do10} Dopita M.~A., Calzetti D., Ma{\'\i}z Apell{\'a}niz J., Blair W.~P., Long K.~S., Mutchler M., Whitmore B.~C., et al., 2010, {\color {navy}Ap\&SS, 330, 123 }

\bibitem[\protect\citeauthoryear{Dubner \& Giacani}{2015}]{Du15}Dubner G., Giacani E., 2015, {\color {navy}A\&ARv, 23, 3}

\bibitem[\protect\citeauthoryear{Ferrand \& Safi-Harb}{2012}]{FeSa12}Ferrand G., Safi-Harb S., 2012, {\color {navy}AdSpR, 49, 1313}

\bibitem[\protect\citeauthoryear{Fesen}{1984}]{Fe84} Fesen R.~A., 1984, {\color {navy}ApJ, 281, 658}

\bibitem[\protect\citeauthoryear{Fesen et al.}{1985}]{Fe85} Fesen R. A., Blair W. P., Kirshner R. P., 1985, ApJ, 292, 29

%\bibitem[\protect\citeauthoryear{Fesen \& Kirshner}{2010}]{FeKi80}Fesen R. A., Kirshner R. P., 1980, {\color {navy}ApJ, 242, 1023}

\bibitem[\protect\citeauthoryear{Fesen \& Milisavljevic}{2010}]{Fe10} Fesen R.~A., Milisavljevic D., 2010, {\color {navy}AJ, 140, 1163} 

\bibitem[\protect\citeauthoryear{Fesen et al.}{2019}]{Fe19} Fesen R. A., Neustadt J. M. M., How T. G., Black C. S. 2019, {\color {navy}MNRAS, 486, 4701}

\bibitem[\protect\citeauthoryear{Fesen et al.}{2020}]{Fe20} Fesen R. A., Weil K. E., Raymond J. C., et al. 2020, {\color {navy}MNRAS, 498, 5194}

\bibitem[\protect\citeauthoryear{Frew \& Parker}{2010}]{Frew2010}Frew D. J., Parker Q. A., 2010, PASA, 27, 129

\bibitem[\protect\citeauthoryear{Green}{2019}]{Gr19}Green D. A., 2019, {\color {navy}JApA, 40, 36}

\bibitem[\protect\citeauthoryear{Jackson, Safi-Harb, \& Kothes}{2014}]{Jackson14} Jackson M.~S., Safi-Harb S., Kothes R., 2014, {\color {navy}MNRAS, 444, 2228}

\bibitem[\protect\citeauthoryear{Kaler}{1976}]{Ka76}Kaler J. B., 1976, {\color {navy} ApJS, 31, 517}

\bibitem[\protect\citeauthoryear{Kothes}{2003}]{Kothes03} Kothes R., 2003, {\color {navy}A\&A, 408, 187}

\bibitem[\protect\citeauthoryear{Mathewson \& Clarke}{1973}]{MaCl73} Mathewson D.~S., Clarke J.~N., 1973, {\color {navy}ApJ, 180, 725} 

\bibitem[\protect\citeauthoryear{Matonick \& Fesen}{1997}]{MaFe97} Matonick D.~M., Fesen R.~A., 1997, {\color {navy}ApJS, 112, 49}

\bibitem[\protect\citeauthoryear{Mavromatakis et al.}{2009}]{Ma09}Mavromatakis F., Boumis, P., Meaburn, J., Caulet A., 2009, {\color {navy} A\&A, 503, 129}

\bibitem[\protect\citeauthoryear{Oke}{1990}]{Ok90} Oke J. B., 1990, {\color {navy}AJ, 99, 1621}

\bibitem[\protect\citeauthoryear{Osterbrock \& Ferland}{2006}]{OsFe06} Osterbrock D.~E., Ferland G.~J., 2006, Astrophysics of Gaseous Nebulae and Active Galactic Nuclei (Mill Valley, CA: Univ. Science Books)

\bibitem[\protect\citeauthoryear{Predehl \& Schmitt}{1995}]{Pr95}Predehl P., Schmitt J. H. M. M., 1995, {\color {navy}A\&A, 293, 889}

\bibitem[\protect\citeauthoryear{Raymond}{1979}]{Ra79} Raymond J. C., 1979, {\color {navy}ApJS, 39, 1}

\bibitem[\protect\citeauthoryear{Raymond  et al.}{1988}]{Ra88} Raymond J. C., Hester J. J., Cox D., Blair W. P., Fesen R. A., Gull T. R., 1988, ApJ, 324, 869


\bibitem[\protect\citeauthoryear{Sezer et al.}{2012}]{Se12} Sezer A., G{\"o}k F., Aktekin E., 2012, {\color {navy}MNRAS, 427, 1168}

\bibitem[\protect\citeauthoryear{Shull \& McKee}{1979}]{Sh79} Shull J. M., McKee C. F., 1979, {\color {navy}ApJ, 227, 131}

\bibitem[\protect\citeauthoryear{Stupar et al.}{2007}]{St07} Stupar M., Parker Q.~A., Filipovi{\'c} M.~D., Frew D.~J., Boji{\v{c}}i{\'c} I., Aschenbach B., 2007, {\color {navy}MNRAS, 381, 377}

\bibitem[\protect\citeauthoryear{Stupar \& Parker}{2009}]{St09} Stupar M., Parker Q.~A., 2009, {\color {navy}MNRAS, 394, 1791}

\bibitem[\protect\citeauthoryear{Stupar \& Parker}{2011}]{St11} Stupar M., Parker Q.~A., 2011, {\color {navy}MNRAS, 414, 2282} 

\bibitem[\protect\citeauthoryear{Taylor et al.}{2003}]{Ta03} Taylor A. R. et al., 2003, {\color {navy}AJ, 125, 3145}


\end{thebibliography}
\end{document}